\documentclass[sigconf]{acmart}
\AtBeginDocument{%
  \providecommand\BibTeX{{%
    \normalfont B\kern-0.5em{\scshape i\kern-0.25em b}\kern-0.8em\TeX}}}

\copyrightyear{2021}
\acmYear{2021}
\setcopyright{rightsretained}
\acmConference[IDC '21]{Interaction Design and Children}{June 24--30, 2021}{Athens, Greece}
\acmBooktitle{Interaction Design and Children (IDC '21), June 24--30, 2021, Athens, Greece}\acmDOI{10.1145/3459990.3465185}
\acmISBN{978-1-4503-8452-0/21/06}

\usepackage{float}

\setcopyright{rightsretained}
\begin{document}

\title{Understanding the Role of Digital Technology in the Transitions of Refugee Families with Young Children into A New Culture: A Case Study of Scotland }

\author{Valentina Andries}
\affiliation{%
  \institution{School of Education, The University of Edinburgh}
  \city{Edinburgh}
  \country{United Kingdom}}
\email{valentina.andries@ed.ac.uk}

\author{Sabina Savadova}
\affiliation{%
  \institution{School of Education, The University of Edinburgh}
  \city{Edinburgh}
  \country{United Kingdom}}
\email{sabina.savadova@ed.ac.uk}


\begin{abstract}
  The worldwide refugee crisis is a major current challenge, affecting the health and education of millions of families with children due to displacement. Despite the various challenges and risks of migration practices, numerous refugee families have access to interactive technologies during these processes. The aim of this ongoing study is to explore the role of technologies in the transitions of refugee families in Scotland. Based on Tudge’s ecocultural theory, a qualitative case-study approach has been adopted. Semi-structured interviews have been conducted with volunteers who work with refugee families in a big city in Scotland, and proxy observations of young children were facilitated remotely by their refugee parents. A preliminary overview of the participants’ insights of the use and role of technology for transitioning into a new culture is provided here.
\end{abstract}



\maketitle

\section{INTRODUCTION}
According to UNICEF, the number of international migrants reached 272 million in 2019, and 33 million of the migrants were displaced children \cite{UNICEF}. “A refugee is defined as an individual who has to flee from his or her home country to escape persecution for different reasons, including race, religion, nationality or political oppression”\cite{UNHCR}. Scotland (United Kingdom) is one of the leading countries welcoming refugees and since 2015, the country has pledged to resettle 20,000 refugees under the UK’s Syrian Resettlement Programme. The challenges associated with displaced children are not new, however they are gaining increased international visibility \cite{menjivar2019undocumented}.

Many families who are displaced go through assimilation processes related to new places, languages, and cultures, facing stressful circumstances as a result \cite{tyrer2014school}. In relation to language acquisition in a new country, families use professional services to receive information and resources about language learning and cultural adaptation \cite{perry2009genres}. Nonetheless, their children can make a significant contribution to their families' transitions into the new country \cite{orellana2003other}. The children spend most of their time engaging in activities that allow for language learning, and adaptation to the new socio-cultural norms quickly and more naturally. 

\subsection{Refugee Families and Technologies}
The increased exposure and reliance of migrants on interactive technologies before and after the relocation process is becoming of interest to the human-computer interaction research community \cite{dekker2018smart, kutscher2018ambivalent, hourcade2019child}. Recent studies such as Kaufmann \cite{kaufmann2018navigating} indicate that smartphones are increasingly being used for everyday language learning and translation \cite{sharples2013mobile} because they can support quick, situated and informal learning \cite{kukulska2015mobile}. In a study with displaced youth from East African, Latin American and Asian countries, Fisher and Yafi \cite{fisher2018syrian} found that young refugees use technology to help their family in various situations, such as to understand and acquire socio-cultural norms, to translate information as well as to navigate geography. 

Nonetheless, with regards to younger refugee children or their caregivers, few technologies are currently adapted to their needs, preferences and abilities \cite{hourcade2019child}. Hourcade et al. \cite{hourcade2019child} discuss that technologies could be designed for specific purposes, particularly in order to facilitate cultural transition and immersion, to support essential communication with family and relatives who are left behind, and to enable their connection with local communities and newly-found peers. 

Although the main purpose of the current study was to understand the refugee families’ use of already existing technologies, requirements for technology design have also emerged qualitatively. These are discussed in a subsequent section (Interviews with Befrienders). Considering all the existing and possible problems refugee families face in  host countries, a number of intervention programs have been put in place to assist families in need, and many of these programmes have been found helpful by all the parties involved \cite{fazel2009school}. While it is not our intention to develop and conduct intervention programmes, the outputs of the project may support refugee families in their transition process, alleviating some of the stress experienced as a result of having to rapidly acquire a new language for cultural adaptation.

\begin{figure*}
  \centering
  \includegraphics[width=\textwidth]{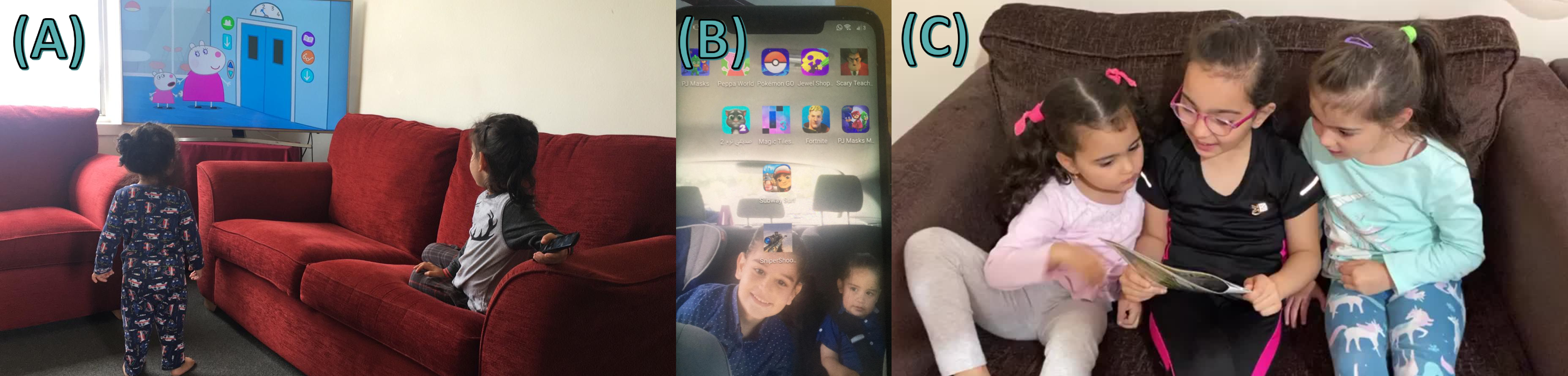}
  \caption{Pictures of children from the two families. Figure (A) depicts the children watching cartoons in English on TV; Figure (B) showcases the app selection on their parents' smartphone, and Figure (C) shows the older child reading stories to the young siblings.}
    ~\label{figure:ChildrenActivities}
\end{figure*}

\subsection{Children and Technology Use}
Over the last decade the number of households owning different kinds of digital media devices has increased rapidly according to the surveys conducted in countries of the Global North \cite{MediaTechnologyMonitor, Ofcom, CommonSenseMedia}. Recent research has revealed that digital technologies are already becoming a part of their everyday practices and children’s exposure to digital devices begins well before school years. Children often start living in a media-rich world through their parents, care-givers and siblings from birth and they are becoming increasingly adept at using digital technologies at younger ages \cite{marsh2011young, o2011young, plowman2012preschool, aubrey2014confidence, flewitt2014touching, marsh2015exploring}. 

Children use digital technologies for a multitude of purposes: to learn, to interact and to understand the surroundings \cite{marsh2011young} or to play games and watch videos \cite{holloway2013zero}. Nevertheless, children’s interactions with digital technologies compose only one part of their daily life in addition to their other everyday activities \cite{plowman2012preschool, wong2015mobile, chaudron2018young}.

Although refugee families use their native language at home, they frequently want to use the host country language as part of the process of settling in and also to help their children to adapt to their new schools \cite{ilic2013home}. However, considering that the majority of parents do not know the language of the host country, particularly mothers who spend more time with their children at home, children often become mediators between their families and the host country \cite{orellana2003other}.  

\textbf{This project aims to locate the role of digital media (digital devices, such as smartphones, tablets, TV and laptops) in this mediation process by:}
\begin{enumerate}
  \item \textbf{understanding volunteer befrienders’ use of technology in supporting the refugees’ transition to the Scottish culture,}
  \item \textbf{determining whether children’s interactions with technologies at home have any impact on families’ transitions into the new culture.}
\end{enumerate}


\section{METHODS \& PARTICIPANTS}
The initial plans for this project entailed conducting face-to-face semi-structured interviews with volunteers about their work with the refugee families, as well as focus groups with different refugee family members in situ while they would take part in various social activities organised by the City Council or by charities from a big city in Scotland for their befriending programme. Furthermore, observations of how the refugees’ children interact with digital technology during these events was also part of the initial study design. 

\subsection{Befrienders}
However, the data collection began in late spring 2020, and the design had to be adapted in the context of the COVID-19 pandemic. Consequently, semi-structured interviews were conducted online with volunteers also known as befrienders (N=4), who regularly work with refugee families from Syria in Scotland, in order to make their transition smoother. 

The befriending programmes facilitated by the Scottish government officially last for 9 months initially, entailing systematic meetings, once a week. The meetings can continue on their own after the 9-month period if a relationship is developed, or it comes to an end; usually, friendships develop as they learn about each others’ cultures. The befriending program was initially created for the refugees to explore the surroundings together with the befriender, however the volunteers adapt to the needs of the families. Their connection fosters learning manifold aspects about life in Scotland, from navigating the social or legal system, to food and leisure. English language practice with adult refugees, as well as with the children are usual activities which these befrienders do with the families, or primarily with the refugee parents. 

\subsection{Families}
So far, data has been collected from two refugee families with young children aged 4-8-years old, by using a method called living journals \cite{savadovaliving}. Through the living journals, parents were assigned as ‘proxy’ \cite{plowman2017revisiting}, and they were sent prompts (\textit{Where is your child? With whom? What is your child doing? Why?}) several days during a week. Parents were asked to reply to these prompts with visuals (pictures or 30-second-long videos) and answers to the questions, as inspired by a study by Plowman and Stevenson \cite{plowman2012preschool}. Parents’ answers together with the visuals they sent were collated to create a personalised journal for the participant children in a paper and digital format. The resulting journals were sent to the families in a paper format, representing research output as well as memorabilia. Then parents were invited to discuss the journal of their own children and that of the other participant's children in semi-structured interviews held online. An interpreter facilitated the conversations. Prior to sharing the children’s journal in a digital format online during the discussions, we obtained the families' consent for this purpose.   

Relying on contextualist ecocultural theory, the living journals method is focused around daily activities, which are grounded in cultural values and norms \cite{tudge2006window, savadovaliving}. The theory sheds lights into children’s day-to-day lives through exploring their activities, interests and interlocutors that they are interacting with on a daily basis. This approach supports the importance of documenting technology use in context within a daily life, without disrupting the participants’ routines via the researchers’ physical presence. 

Thematic analysis  has been employed to analyse the data from the volunteers as well as the resulting living journals \cite{braun2013successful}. Each family was treated as a case, and initially their data, together with the visuals, were coded and analysed inductively following our research aim, and then across cases \cite{merriam2015qualitative}. The data analysis was focused on revealing insights into children’s daily lives as first communicated by the befriender interviews, further visualised by the parents in the journals, and discussed in the living journal interviews. 

Ethical approval was granted from the University for this study. Special care was taken to obtain parents and children’s consent and assent to participate in this study, and use their materials in publications and conference presentations. All of the data collection has been carried out remotely, respecting social distancing rules imposed by the COVID-19 pandemic, and allowing for flexibility in data collection. The living journals approach allows the families to observe their children using digital technology in their own home environments, as well as to communicate with the researchers at times that suit them. 

Because of the remote research approach, the researchers’ involvement has been minimised for the study component involving the refugee families. The living journals data was collected by using the instant messaging application WhatsApp, which has the benefit of encryption features. The parents were more confident to use WhatsApp to send and receive data because they regularly use that for communication with befrienders and their families in their home countries, instead of platforms such as Collaborate or Microsoft Teams.

\section{FINDINGS}
\subsection{Interviews with Befrienders}
Thus far, the interviews with the befrienders (N=4) have revealed a few themes. These are discussed below, as well as in relation to the themes emerging from the families’ data.

\subsubsection{Technology Access}
The volunteers made observations previously reinforced in the literature, regarding the refugees' access to and use of smartphone technology. Most families that they have interacted with have smartphones; they are seen as a necessity, not a luxury. The refugees need smartphones primarily to communicate with social workers, volunteers, the local authorities and their families; they seem to have appointments all the time with the authorities through their social workers.
\subsubsection{Communication}
According to the befrienders, the refugees seem to have an overall preference for WhatsApp for communicating with each other, for keeping in touch with other refugees in the same city, as well as for talking to their family members and relatives who were left behind. So the befrienders also use WhatsApp to set meetings and communicate with the refugees. If the befrienders happen to only have limited Arabic language skills, or the nature of the discussion between them and the refugee adults implies more complexity, the latter may involve their children or grandchildren in the conversation. The children may have more developed language skills and they feel proud of their role as enablers.

\subsubsection{WhatsApp for Homework}
Due to the pandemic, the befrienders decided (by discussing in their own WhatsApp group) to help the refugees’ children however needed for learning and practising English. They made this decision due to closing of schools and the uncertainty about socialising. This approach proved to be particularly valuable for families with children who arrived in Scotland shortly before the pandemic affected all essential activities in the UK in spring 2020. The befrienders explained that they were focusing on helping the children out with the homework online on WhatsApp, such as reading practice and pronunciation. 
\subsubsection{Google Translate}
According to the befrienders, the refugees are aware of various apps and dictionaries that they can use to communicate. English language classes are also facilitated through the City Council for the adults that are interested or available for that. However, they reported that they mainly make use of Google for acquiring information, and Google translate specifically for language purposes. Nonetheless, they note that, while Google translate can be useful for some very basic communication, its use is rather limited beyond that. 

The befrienders interestingly point out that communication can quickly be impeded when using such apps due to a lack of dialectic input/output. More specifically, issues with google translate can refer to the system using standard Arabic, whereas Syrians speak levantine Arabic, which is the spoken language. Thus, different nuances of communication are lost in translation, resulting in potentially awkward moments. One of the befrienders discusses that “it messes things up even more”, leading to broken communication as it is not easy to identify the correct synonyms or nuances to use depending on the type of conversation that they engage in. 

Another befriender emphasised that the National Health Services should provide interpreters for the refugees’ visits to the doctor but in practice, that is not always possible. This is just one example of situations where interactive technology could be very useful to convey messages efficiently. 

\subsubsection{Children as Language Enablers}
Nonetheless, the befrienders describe themselves and the adult refugees as not very technological, as opposed to the children. They discuss that, as expected, the children learn English faster than the parents, and that “some people learn to depend on their children” for communication purposes. In some cases, the children who get to speak English very well would playfully correct their parents, translate phrases for them, or help the befrienders to communicate with their parents when Google translate fails. 

\subsubsection{Early Design Requirements for A Customisable Language Tool}
For communicating with people with very limited English, the volunteers emphasise their reliance on visual materials such as picture dictionaries, flashcards which the befrienders would make specifically for themes relevant for the different families that they work with, or Google images to refer to specific objects. The volunteers describe turning such activities into a language exchange, the former learning words in Arabic and the refugees learning something in English, while the children help as well. 

The befrienders discuss that they would find it beneficial to use a digital tool for language exchange. More specifically, such a tool should allow the users to customise its content according to themes or family needs. Children’s homework could be one of these themes, as emerging from the interviews, which could then be further subdivided into topics. One of the befrienders reflects that such a tool could be used instead of flashcards, to which videos could be added as well, to facilitate memorising. Another befriender discusses that a visual dictionary could be built within such a platform, and these could be envisaged as family activities with the help of the volunteers. They also reflect on further personalising such a tool, by allowing the users to build a useful phrase bank that would appropriately reflect their dialect, and to express different nuances in communication.

\subsection{Living Journals and Interviews with Families}
The living journals from the families reinforced some of the themes presented earlier, such as the families using smartphones and WhatsApp to communicate with their families from overseas. The pictures sent by the parents, as prompted by the researchers (\textit{What are they [children] doing now?}) revealed that the children use digital technologies as part of their daily activities, alongside arts and crafts, books, exercise books for homework and outdoors activities. Specifically to language acquisition, the living journals also showcased the children engaging in various activities which facilitate this process. Those ranged from doing English homework with a view to support more formal literacy practices, as well as by watching cartoons in English on TV (Figure A), listening to nursery rhymes, or using different apps in English on their tablets, and the parents’ smartphones (Figure B). 

From the volunteers' accounts as well as from the living journals, storytelling emerged as an activity that also facilitates language learning. Furthermore, storytelling was used to preserve their native language as well. As it can be noted in Figure C, the older child from one of the families that has taken part in the study so far, is being portrayed reading a story in Arabic to their younger sisters. The parents and the volunteers want to encourage communication in their native language to their children so as to keep it alive. They also discussed that there used to be volunteer groups that organise storytelling times in both Arabic and English, which they enjoyed attending prior to the pandemic.

\section{CONCLUSION}
To summarise these preliminary findings, it is apparent that interactive technologies play an important part in refugee families' transitions into the Scottish culture. Exploring refugee children’s interactions with digital technologies within their daily lives through the lens of ecocultural theory allowed us to reveal the ways in which children contributed to the families’ day-to-day lives and their adaptation in the new country. However, it is important to further unravel the different needs and preferences of refugee families and volunteers regarding language acquisition and communication, and how digital technologies could meet them.

At this stage, as recruitment is still ongoing, it is difficult to draw definite conclusions regarding the design of a digital platform for communication, from the perspective of the refugee families' and that of the young children. More comprehensive reflections about the opportunities afforded by existing technologies, as well as their shortcomings when it comes to being used in ways that were not initially intended shall be provided in further work, complemented by more data. Nevertheless, the role of children in their parents' language acquisition, as well as the guiding and facilitating influence of the befrienders has emerged from the current data. The purpose of the research is to ultimately compile a guidebook for the families and the volunteers, highlighting how digital technology can support language acquisition for cultural assimilation. Lastly, design requirements can be provided, with a view to developing an English language-facilitating tool that would support the families' transition. 

\section{Acknowledgments}
We would like to extend our gratitude to the befrienders and refugee families who have kindly agreed to take part in our research. We thank the Universitas21 network for granting us in 2019 a Graduate Collaborative Research Award of 5000USD, Professor Lydia Plowman for her invaluable guidance, as well as Alexandra Pennycuick and Laura Iosip.

\bibliographystyle{ACM-Reference-Format}
\bibliography{references}


\begin{thebibliography}{32}


\ifx \showCODEN    \undefined \def \showCODEN     #1{\unskip}     \fi
\ifx \showDOI      \undefined \def \showDOI       #1{#1}\fi
\ifx \showISBNx    \undefined \def \showISBNx     #1{\unskip}     \fi
\ifx \showISBNxiii \undefined \def \showISBNxiii  #1{\unskip}     \fi
\ifx \showISSN     \undefined \def \showISSN      #1{\unskip}     \fi
\ifx \showLCCN     \undefined \def \showLCCN      #1{\unskip}     \fi
\ifx \shownote     \undefined \def \shownote      #1{#1}          \fi
\ifx \showarticletitle \undefined \def \showarticletitle #1{#1}   \fi
\ifx \showURL      \undefined \def \showURL       {\relax}        \fi
\providecommand\bibfield[2]{#2}
\providecommand\bibinfo[2]{#2}
\providecommand\natexlab[1]{#1}
\providecommand\showeprint[2][]{arXiv:#2}

\bibitem[\protect\citeauthoryear{??}{UNH}{1992}]%
        {UNHCR}
 \bibinfo{year}{1992}\natexlab{}.
\newblock \bibinfo{title}{Handbook on Procedures and Criteria for Determining
  Refugee Status under the 1951 Convention and the 1967 Protocol relating to
  the Status of Refugees}.
\newblock
\newblock


\bibitem[\protect\citeauthoryear{??}{Com}{2018}]%
        {CommonSenseMedia}
 \bibinfo{year}{2018}\natexlab{}.
\newblock \bibinfo{title}{Common Sense Census: Media Use by Kids Age Zero to
  Eight 2017}.
\newblock
\newblock


\bibitem[\protect\citeauthoryear{??}{Med}{2018}]%
        {MediaTechnologyMonitor}
 \bibinfo{year}{2018}\natexlab{}.
\newblock \bibinfo{title}{Media Technology Monitor Report}.
\newblock
\newblock


\bibitem[\protect\citeauthoryear{??}{Ofc}{2019}]%
        {Ofcom}
 \bibinfo{year}{2019}\natexlab{}.
\newblock \bibinfo{title}{Communications Market Report}.
\newblock
\newblock


\bibitem[\protect\citeauthoryear{??}{UNI}{2020}]%
        {UNICEF}
 \bibinfo{year}{2020}\natexlab{}.
\newblock \bibinfo{title}{UNICEF Report on Child migration April 2020}.
\newblock
\newblock


\bibitem[\protect\citeauthoryear{Aubrey and Dahl}{Aubrey and Dahl}{2014}]%
        {aubrey2014confidence}
\bibfield{author}{\bibinfo{person}{Carol Aubrey} {and} \bibinfo{person}{Sarah
  Dahl}.} \bibinfo{year}{2014}\natexlab{}.
\newblock \showarticletitle{The confidence and competence in information and
  communication technologies of practitioners, parents and young children in
  the Early Years Foundation Stage}.
\newblock \bibinfo{journal}{\emph{Early years}} \bibinfo{volume}{34},
  \bibinfo{number}{1} (\bibinfo{year}{2014}), \bibinfo{pages}{94--108}.
\newblock


\bibitem[\protect\citeauthoryear{Braun and Clarke}{Braun and Clarke}{2013}]%
        {braun2013successful}
\bibfield{author}{\bibinfo{person}{Virginia Braun} {and}
  \bibinfo{person}{Victoria Clarke}.} \bibinfo{year}{2013}\natexlab{}.
\newblock \bibinfo{booktitle}{\emph{Successful qualitative research: A
  practical guide for beginners}}.
\newblock \bibinfo{publisher}{sage}.
\newblock


\bibitem[\protect\citeauthoryear{Chaudron, Di~Gioia, and Gemo}{Chaudron
  et~al\mbox{.}}{2018}]%
        {chaudron2018young}
\bibfield{author}{\bibinfo{person}{Stephane Chaudron}, \bibinfo{person}{Rosanna
  Di~Gioia}, {and} \bibinfo{person}{Monica Gemo}.}
  \bibinfo{year}{2018}\natexlab{}.
\newblock \showarticletitle{Young children (0-8) and digital technology, a
  qualitative study across Europe}.
\newblock \bibinfo{journal}{\emph{JRC Science for Policy Report}}
  (\bibinfo{year}{2018}).
\newblock


\bibitem[\protect\citeauthoryear{Dekker, Engbersen, Klaver, and Vonk}{Dekker
  et~al\mbox{.}}{2018}]%
        {dekker2018smart}
\bibfield{author}{\bibinfo{person}{Rianne Dekker}, \bibinfo{person}{Godfried
  Engbersen}, \bibinfo{person}{Jeanine Klaver}, {and} \bibinfo{person}{Hanna
  Vonk}.} \bibinfo{year}{2018}\natexlab{}.
\newblock \showarticletitle{Smart refugees: How Syrian asylum migrants use
  social media information in migration decision-making}.
\newblock \bibinfo{journal}{\emph{Social Media+ Society}} \bibinfo{volume}{4},
  \bibinfo{number}{1} (\bibinfo{year}{2018}),
  \bibinfo{pages}{2056305118764439}.
\newblock


\bibitem[\protect\citeauthoryear{Fazel, Doll, and Stein}{Fazel
  et~al\mbox{.}}{2009}]%
        {fazel2009school}
\bibfield{author}{\bibinfo{person}{Mina Fazel}, \bibinfo{person}{Helen Doll},
  {and} \bibinfo{person}{Alan Stein}.} \bibinfo{year}{2009}\natexlab{}.
\newblock \showarticletitle{A school-based mental health intervention for
  refugee children: An exploratory study}.
\newblock \bibinfo{journal}{\emph{Clinical Child Psychology and Psychiatry}}
  \bibinfo{volume}{14}, \bibinfo{number}{2} (\bibinfo{year}{2009}),
  \bibinfo{pages}{297--309}.
\newblock


\bibitem[\protect\citeauthoryear{Fisher and Yafi}{Fisher and Yafi}{2018}]%
        {fisher2018syrian}
\bibfield{author}{\bibinfo{person}{Karen~E Fisher} {and} \bibinfo{person}{Eiad
  Yafi}.} \bibinfo{year}{2018}\natexlab{}.
\newblock \showarticletitle{Syrian Youth in Za'atari Refugee Camp as ICT
  Wayfarers: An Exploratory Study Using LEGO and Storytelling}. In
  \bibinfo{booktitle}{\emph{Proceedings of the 1st ACM SIGCAS Conference on
  Computing and Sustainable Societies}}. \bibinfo{pages}{1--12}.
\newblock


\bibitem[\protect\citeauthoryear{Flewitt, Kucirkova, and Messer}{Flewitt
  et~al\mbox{.}}{2014}]%
        {flewitt2014touching}
\bibfield{author}{\bibinfo{person}{Rosie Flewitt}, \bibinfo{person}{Natalia
  Kucirkova}, {and} \bibinfo{person}{David Messer}.}
  \bibinfo{year}{2014}\natexlab{}.
\newblock \showarticletitle{Touching the virtual, touching the real: iPads and
  enabling literacy for students experiencing disability}.
\newblock \bibinfo{journal}{\emph{Australian Journal of Language \& Literacy}}
  \bibinfo{volume}{37}, \bibinfo{number}{2} (\bibinfo{year}{2014}),
  \bibinfo{pages}{107--116}.
\newblock


\bibitem[\protect\citeauthoryear{Holloway, Green, and Livingstone}{Holloway
  et~al\mbox{.}}{2013}]%
        {holloway2013zero}
\bibfield{author}{\bibinfo{person}{Donell Holloway}, \bibinfo{person}{Lelia
  Green}, {and} \bibinfo{person}{Sonia Livingstone}.}
  \bibinfo{year}{2013}\natexlab{}.
\newblock \showarticletitle{Zero to eight: Young children and their internet
  use}.
\newblock  (\bibinfo{year}{2013}).
\newblock


\bibitem[\protect\citeauthoryear{Hourcade, Antle, Giannakos, Fails, Read,
  Markopoulos, Garzotto, and Palumbos}{Hourcade et~al\mbox{.}}{2019}]%
        {hourcade2019child}
\bibfield{author}{\bibinfo{person}{Juan~Pablo Hourcade},
  \bibinfo{person}{Alissa~N Antle}, \bibinfo{person}{Michail Giannakos},
  \bibinfo{person}{Jerry~Alan Fails}, \bibinfo{person}{Janet~C Read},
  \bibinfo{person}{Panos Markopoulos}, \bibinfo{person}{Franca Garzotto}, {and}
  \bibinfo{person}{Andrea Palumbos}.} \bibinfo{year}{2019}\natexlab{}.
\newblock \showarticletitle{Child-computer interaction sig: Designing for
  refugee children}. In \bibinfo{booktitle}{\emph{Extended Abstracts of the
  2019 CHI Conference on Human Factors in Computing Systems}}.
  \bibinfo{pages}{1--4}.
\newblock


\bibitem[\protect\citeauthoryear{Ili{\'c}}{Ili{\'c}}{2013}]%
        {ilic2013home}
\bibfield{author}{\bibinfo{person}{Vesna Ili{\'c}}.}
  \bibinfo{year}{2013}\natexlab{}.
\newblock \showarticletitle{Home-literacy practices and academic language
  skills of migrant pupils}.
\newblock \bibinfo{journal}{\emph{Tertium comparationis}} \bibinfo{volume}{18},
  \bibinfo{number}{2} (\bibinfo{year}{2013}), \bibinfo{pages}{190--208}.
\newblock


\bibitem[\protect\citeauthoryear{Kaufmann}{Kaufmann}{2018}]%
        {kaufmann2018navigating}
\bibfield{author}{\bibinfo{person}{Katja Kaufmann}.}
  \bibinfo{year}{2018}\natexlab{}.
\newblock \showarticletitle{Navigating a new life: Syrian refugees and their
  smartphones in Vienna}.
\newblock \bibinfo{journal}{\emph{Information, Communication \& Society}}
  \bibinfo{volume}{21}, \bibinfo{number}{6} (\bibinfo{year}{2018}),
  \bibinfo{pages}{882--898}.
\newblock


\bibitem[\protect\citeauthoryear{Kukulska-Hulme, Gaved, Paletta, Scanlon,
  Jones, and Brasher}{Kukulska-Hulme et~al\mbox{.}}{2015}]%
        {kukulska2015mobile}
\bibfield{author}{\bibinfo{person}{Agnes Kukulska-Hulme}, \bibinfo{person}{Mark
  Gaved}, \bibinfo{person}{Lucas Paletta}, \bibinfo{person}{Eileen Scanlon},
  \bibinfo{person}{Ann Jones}, {and} \bibinfo{person}{Andrew Brasher}.}
  \bibinfo{year}{2015}\natexlab{}.
\newblock \showarticletitle{Mobile incidental learning to support the inclusion
  of recent immigrants}.
\newblock \bibinfo{journal}{\emph{Ubiquitous Learning: an international
  journal}} \bibinfo{volume}{7}, \bibinfo{number}{2} (\bibinfo{year}{2015}),
  \bibinfo{pages}{9--21}.
\newblock


\bibitem[\protect\citeauthoryear{Kutscher and Kre{\ss}}{Kutscher and
  Kre{\ss}}{2018}]%
        {kutscher2018ambivalent}
\bibfield{author}{\bibinfo{person}{Nadia Kutscher} {and}
  \bibinfo{person}{Lisa-Marie Kre{\ss}}.} \bibinfo{year}{2018}\natexlab{}.
\newblock \showarticletitle{The ambivalent potentials of social media use by
  unaccompanied minor refugees}.
\newblock \bibinfo{journal}{\emph{Social Media+ Society}} \bibinfo{volume}{4},
  \bibinfo{number}{1} (\bibinfo{year}{2018}),
  \bibinfo{pages}{2056305118764438}.
\newblock


\bibitem[\protect\citeauthoryear{Marsh}{Marsh}{2011}]%
        {marsh2011young}
\bibfield{author}{\bibinfo{person}{Jackie Marsh}.}
  \bibinfo{year}{2011}\natexlab{}.
\newblock \showarticletitle{Young children's literacy practices in a virtual
  world: Establishing an online interaction order}.
\newblock \bibinfo{journal}{\emph{Reading research quarterly}}
  \bibinfo{volume}{46}, \bibinfo{number}{2} (\bibinfo{year}{2011}),
  \bibinfo{pages}{101--118}.
\newblock


\bibitem[\protect\citeauthoryear{Marsh, Plowman, Yamada-Rice, Bishop, Lahmar,
  Scott, Davenport, Davis, French, Piras, et~al\mbox{.}}{Marsh
  et~al\mbox{.}}{2015}]%
        {marsh2015exploring}
\bibfield{author}{\bibinfo{person}{Jackie Marsh}, \bibinfo{person}{Lydia
  Plowman}, \bibinfo{person}{Dylan Yamada-Rice}, \bibinfo{person}{Julia
  Bishop}, \bibinfo{person}{Jamel Lahmar}, \bibinfo{person}{Fiona Scott},
  \bibinfo{person}{Andrew Davenport}, \bibinfo{person}{Simon Davis},
  \bibinfo{person}{Katie French}, \bibinfo{person}{Maddalena Piras},
  {et~al\mbox{.}}} \bibinfo{year}{2015}\natexlab{}.
\newblock \showarticletitle{Exploring Play and Creativity in Pre-schooler's use
  of apps: Final Project Report}.
\newblock  (\bibinfo{year}{2015}).
\newblock


\bibitem[\protect\citeauthoryear{Menj{\'\i}var and Perreira}{Menj{\'\i}var and
  Perreira}{2019}]%
        {menjivar2019undocumented}
\bibfield{author}{\bibinfo{person}{Cecilia Menj{\'\i}var} {and}
  \bibinfo{person}{Krista~M Perreira}.} \bibinfo{year}{2019}\natexlab{}.
\newblock \bibinfo{title}{Undocumented and unaccompanied: children of migration
  in the European Union and the United States}.
\newblock
\newblock


\bibitem[\protect\citeauthoryear{Merriam and Tisdell}{Merriam and
  Tisdell}{2015}]%
        {merriam2015qualitative}
\bibfield{author}{\bibinfo{person}{Sharan~B Merriam} {and}
  \bibinfo{person}{Elizabeth~J Tisdell}.} \bibinfo{year}{2015}\natexlab{}.
\newblock \bibinfo{booktitle}{\emph{Qualitative research: A guide to design and
  implementation}}.
\newblock \bibinfo{publisher}{John Wiley \& Sons}.
\newblock


\bibitem[\protect\citeauthoryear{O'Hara}{O'Hara}{2011}]%
        {o2011young}
\bibfield{author}{\bibinfo{person}{Mark O'Hara}.}
  \bibinfo{year}{2011}\natexlab{}.
\newblock \showarticletitle{Young children’s ICT experiences in the home:
  Some parental perspectives}.
\newblock \bibinfo{journal}{\emph{Journal of Early Childhood Research}}
  \bibinfo{volume}{9}, \bibinfo{number}{3} (\bibinfo{year}{2011}),
  \bibinfo{pages}{220--231}.
\newblock


\bibitem[\protect\citeauthoryear{Orellana, Reynolds, Dorner, and Meza}{Orellana
  et~al\mbox{.}}{2003}]%
        {orellana2003other}
\bibfield{author}{\bibinfo{person}{Marjorie~Faulstich Orellana},
  \bibinfo{person}{Jennifer Reynolds}, \bibinfo{person}{Lisa Dorner}, {and}
  \bibinfo{person}{Mar{\'\i}a Meza}.} \bibinfo{year}{2003}\natexlab{}.
\newblock \showarticletitle{In other words: Translating or “para-phrasing”
  as a family literacy practice in immigrant households}.
\newblock \bibinfo{journal}{\emph{Reading research quarterly}}
  \bibinfo{volume}{38}, \bibinfo{number}{1} (\bibinfo{year}{2003}),
  \bibinfo{pages}{12--34}.
\newblock


\bibitem[\protect\citeauthoryear{Perry}{Perry}{2009}]%
        {perry2009genres}
\bibfield{author}{\bibinfo{person}{Kristen~H Perry}.}
  \bibinfo{year}{2009}\natexlab{}.
\newblock \showarticletitle{Genres, contexts, and literacy practices: Literacy
  brokering among Sudanese refugee families}.
\newblock \bibinfo{journal}{\emph{Reading Research Quarterly}}
  \bibinfo{volume}{44}, \bibinfo{number}{3} (\bibinfo{year}{2009}),
  \bibinfo{pages}{256--276}.
\newblock


\bibitem[\protect\citeauthoryear{Plowman}{Plowman}{2017}]%
        {plowman2017revisiting}
\bibfield{author}{\bibinfo{person}{Lydia Plowman}.}
  \bibinfo{year}{2017}\natexlab{}.
\newblock \showarticletitle{Revisiting ethnography by proxy}.
\newblock \bibinfo{journal}{\emph{International Journal of Social Research
  Methodology}} \bibinfo{volume}{20}, \bibinfo{number}{5}
  (\bibinfo{year}{2017}), \bibinfo{pages}{443--454}.
\newblock


\bibitem[\protect\citeauthoryear{Plowman, Stevenson, Stephen, and
  McPake}{Plowman et~al\mbox{.}}{2012}]%
        {plowman2012preschool}
\bibfield{author}{\bibinfo{person}{Lydia Plowman}, \bibinfo{person}{Olivia
  Stevenson}, \bibinfo{person}{Christine Stephen}, {and}
  \bibinfo{person}{Joanna McPake}.} \bibinfo{year}{2012}\natexlab{}.
\newblock \showarticletitle{Preschool children’s learning with technology at
  home}.
\newblock \bibinfo{journal}{\emph{Computers \& Education}}
  \bibinfo{volume}{59}, \bibinfo{number}{1} (\bibinfo{year}{2012}),
  \bibinfo{pages}{30--37}.
\newblock


\bibitem[\protect\citeauthoryear{Savadova and Plowman}{Savadova and
  Plowman}{2020}]%
        {savadovaliving}
\bibfield{author}{\bibinfo{person}{Sabina Savadova} {and}
  \bibinfo{person}{Lydia Plowman}.} \bibinfo{year}{2020}\natexlab{}.
\newblock \showarticletitle{Living journals: Families interpreting young
  children’s everyday lives in Azerbaijan}. In \bibinfo{booktitle}{\emph{2020
  annual meeting of the American Educational Research Association (AERA), San
  Francisco, CA, United States (Conference cancelled)}}.
\newblock


\bibitem[\protect\citeauthoryear{Sharples}{Sharples}{2013}]%
        {sharples2013mobile}
\bibfield{author}{\bibinfo{person}{Mike Sharples}.}
  \bibinfo{year}{2013}\natexlab{}.
\newblock \showarticletitle{Mobile learning: research, practice and
  challenges}.
\newblock \bibinfo{journal}{\emph{Distance Education in China}}
  \bibinfo{volume}{3}, \bibinfo{number}{5} (\bibinfo{year}{2013}),
  \bibinfo{pages}{5--11}.
\newblock


\bibitem[\protect\citeauthoryear{Tudge, Doucet, Odero, Sperb, Piccinini, and
  Lopes}{Tudge et~al\mbox{.}}{2006}]%
        {tudge2006window}
\bibfield{author}{\bibinfo{person}{Jonathan~RH Tudge},
  \bibinfo{person}{Fabienne Doucet}, \bibinfo{person}{Dolphine Odero},
  \bibinfo{person}{Tania~M Sperb}, \bibinfo{person}{Cesar~A Piccinini}, {and}
  \bibinfo{person}{Rita~S Lopes}.} \bibinfo{year}{2006}\natexlab{}.
\newblock \showarticletitle{A window into different cultural worlds: Young
  children's everyday activities in the United States, Brazil, and Kenya}.
\newblock \bibinfo{journal}{\emph{Child development}} \bibinfo{volume}{77},
  \bibinfo{number}{5} (\bibinfo{year}{2006}), \bibinfo{pages}{1446--1469}.
\newblock


\bibitem[\protect\citeauthoryear{Tyrer and Fazel}{Tyrer and Fazel}{2014}]%
        {tyrer2014school}
\bibfield{author}{\bibinfo{person}{Rebecca~A Tyrer} {and} \bibinfo{person}{Mina
  Fazel}.} \bibinfo{year}{2014}\natexlab{}.
\newblock \showarticletitle{School and community-based interventions for
  refugee and asylum seeking children: a systematic review}.
\newblock \bibinfo{journal}{\emph{PloS one}} \bibinfo{volume}{9},
  \bibinfo{number}{2} (\bibinfo{year}{2014}), \bibinfo{pages}{e89359}.
\newblock


\bibitem[\protect\citeauthoryear{Wong}{Wong}{2015}]%
        {wong2015mobile}
\bibfield{author}{\bibinfo{person}{Suzanna So-Har Wong}.}
  \bibinfo{year}{2015}\natexlab{}.
\newblock \showarticletitle{Mobile digital devices and preschoolers’ home
  multiliteracy practices}.
\newblock \bibinfo{journal}{\emph{Language and Literacy}} \bibinfo{volume}{17},
  \bibinfo{number}{2} (\bibinfo{year}{2015}), \bibinfo{pages}{75--90}.
\newblock


\end{thebibliography}
\end{document}